\documentclass[twocolumn,aps,prl,amsmath,amssymb,preprintnumbers]{revtex4}
\begin{document}

\title{Small magnetic charges and monopoles in non-associative quantum mechanics}

\author{Martin Bojowald,$^1$ Suddhasattwa Brahma,$^2$ Umut
  B\"{u}y\"{u}k\c{c}am,$^1$ 
  Jonathan Guglielmon,$^1$ Martijn van Kuppeveld$^1$}
\email{bojowald@gravity.psu.edu, suddhasattwa.brahma@gmail.com,
  ubuyukcam@gmail.com, jag585@psu.edu, mtvankuppeveld@gmail.com}
\affiliation{$^1$ Department of Physics, The Pennsylvania State University, 104
  Davey Lab, University Park, PA 16802, USA\\
$^2$ Asia Pacific Center for Theoretical Physics, Pohang 37673, South
  Korea} 

\begin{abstract}
  Weak magnetic monopoles with a continuum of charges less than the minimum
  implied by Dirac's quantization condition may be possible in non-associative
  quantum mechanics. If a weakly magnetically charged proton in a hydrogen
  atom perturbs the standard energy spectrum only slightly, magnetic charges
  could have escaped detection. Testing this hypothesis requires entirely new
  methods to compute energy spectra in non-associative quantum mechanics. Such
  methods are presented here, and evaluated for upper bounds on the magnetic
  charge of elementary particles.
\end{abstract}

\maketitle

In 1931, Dirac \cite{DiracMonopoles} showed that magnetic monopoles with
charge $g$ can be consistently described by wave functions provided the
quantization condition $eg=N \hbar$ holds with half-integer $N$. Since the
elementary electric charge $e$ (or, rather, the fine structure constant) is
small, the elementary magnetic charge is large. Therefore, there are strict
limits on the possible magnetic charge of, say, a proton in a hydrogen nucleus
because the strong magnetic charge would significantly alter the energy
spectrum \cite{MonopoleHydro}.

The aim of this letter is to point out and analyze the fact that Dirac's
argument relies on properties of wave functions in a Hilbert space, and
therefore implicitly assumes that quantum mechanics is associative. If the
assumption of associativity is dropped, there is no Hilbert-space
representation of the algebra of observables (which by necessicity would
always be associative), but quantum mechanics 
may still be meaningful
\cite{Malcev,JackiwMon,Jackiw,OctQM}. Indeed, the existence of consistent
non-associative algebras for magnetic charge densities has recently been
demonstrated \cite{NonGeoNonAss,MSS1,BakasLuest,MSS2,MSS3}. 
Non-associative quantum mechanics can therefore be defined by replacing the
operator product of observables with an abstract product, such that
$\hat{a}_1(\hat{a}_2\hat{a}_3)\not=(\hat{a}_1\hat{a}_2)\hat{a}_3$ in
general. States are defined as expectation-value functionals that assign
complex numbers $\langle \hat{a}\rangle$ to algebra elements $\hat{a}$,
subject to certain consistency conditions which make sure that uncertainty
relations are respected. No wave functions appear in this formalism, and there
is no analog of ``single-valuedness'' used crucially by Dirac.
Without wave functions, Dirac's argument therefore loses its footing.
Magnetic monopoles are then possible with small charges much less than the
smallest non-zero value, $g_0=\frac{1}{2}\hbar/e$, allowed by Dirac. It is
conceivable that a small magnetic charge of the proton could have escaped
detection in precision spectroscopy such as \cite{HydrogenSpec}.

Here, we show that even a small magnetic charge of the nucleus would
significantly shift the ground-state energy of a hydrogen atom.  To the best
of our knowledge, this is the first time that properties of energy spectra
have been computed in non-associative quantum mechanics. We provide new
methods to compute spectra in an algebraic manner, which may also be useful in
other contexts.

{\em Harmonic oscillator:} We first demonstrate the new methods in an
application to the harmonic oscillator in standard, associative quantum
mechanics.  We have two distinguished observables $\hat{q}$ and $\hat{p}$ with
$[\hat{q},\hat{p}]=i\hbar$, and the quantum Hamiltonian
$\hat{H}=\frac{1}{2}(\hat{p}^2/m+m\omega^2\hat{q}^2)$.

An eigenstate $|\psi_E\rangle$ of $\hat{H}$ with eigenvalue $E$ obeys the
equation $\hat{H}|\psi_E\rangle=E|\psi_E\rangle$, which implies
\begin{equation} \label{eigen}
 \langle\hat{a}(\hat{H}-E)\rangle_E=0
\end{equation}
for the expectation value $\langle\cdot\rangle_E$ taken in $|\psi_E\rangle$,
where $\hat{a}$ can be any polynomial in $\hat{q}$ and $\hat{p}$. We will
first show that (\ref{eigen}), which amounts to infinitely many equations
given the freedom of choosing $\hat{a}$, allows one to compute the spectrum of
$\hat{H}$ even if the eigenstates $|\psi_E\rangle$ are not known.  In
\cite{NonAss,NonAssEffPot}, it has been shown how observables can be computed
using algebraic relations between moments of a state. The methods used here
are closely related to these papers but provide a new application to
energy spectra.  In this way, we will set up a method to compute eigenvalues
without using wave functions or boundary conditions.  The same method can then
be applied to the Coulomb problem in non-associative quantum mechanics.

The demonstration is based on recurrence with respect to the degree of the
polynomial $\hat{a}$ in $\hat{q}$ and $\hat{p}$. The ground-state energy can
be obtained by elementary calculations as follows: First, $\hat{a}=\hat{1}$
(the identity operator) gives
$E=\frac{1}{2}(\langle\hat{p}^2\rangle_E/m+m\omega^2
\langle\hat{q}^2\rangle_E)$. For $\hat{a}$ not the identity, it is useful to
refer to the equation
\begin{equation} \label{aH}
 \langle[\hat{a},\hat{H}]\rangle_E=\langle\hat{a}\hat{H}\rangle_E-
 \overline{\langle\hat{a}^{\dagger}\hat{H}\rangle_E}=
 E\left(\langle\hat{a}\rangle_E-
\overline{\langle\hat{a}^{\dagger}\rangle_E}\right)=0
\end{equation}
with the complex conjugate $\bar{z}$ of a complex number $z$.
In particular, $\langle[\hat{q},\hat{H}]\rangle_E=
i\hbar\langle\hat{p}\rangle_E/m=0$ from $\hat{a}=\hat{q}$ and
$\langle[\hat{p},\hat{H}]\rangle_E= -i\hbar
m\omega^2\langle\hat{q}\rangle_E=0$ from $\hat{a}=\hat{p}$. From
quadratic monomials, we obtain $\langle[\hat{q}^2,\hat{H}]\rangle_E=
\frac{1}{2}i\hbar \langle\hat{q}\hat{p}+\hat{p}\hat{q}\rangle_E/m=0$ and
$\langle[\hat{q}\hat{p},\hat{H}]\rangle_E=
i\hbar(\langle\hat{p}^2\rangle_E/m-m\omega^2
\langle\hat{q}^2\rangle_E)=0$. Therefore, any eigenstate has fluctuations
obeying $\Delta_E p=m\omega\Delta_E q$, and zero covariance
$0=C^E_{qp}=\frac{1}{2}\langle\hat{q}\hat{p}+\hat{p}\hat{q}\rangle_E-
\langle\hat{q}\rangle_E\langle\hat{p}\rangle_E$. From the condition for
$\hat{a}=\hat{1}$, $(\Delta_E q)^2=E/(m\omega^2)$ and $(\Delta_E
p)^2=mE$.

So far, we have computed moments of a bound state in terms of its energy
value $E$. We obtain a restriction on $E$ by making sure that the fluctuations
we derived obey the uncertainty relation:
\begin{equation}
 (\Delta_E q)^2(\Delta_E p)^2-(C^E_{qp})^2= \frac{E^2}{\omega^2}\geq
 \frac{\hbar^2}{4}
\end{equation}
and therefore $E\geq \frac{1}{2} \hbar\omega$. 

In order to evaluate all the conditions imposed on eigenstates by
(\ref{eigen}), we follow \cite{PositiveWigner,MomentsUncert} and introduce the
operators $\hat{T}_{m,n} := (\hat{q}^m\hat{p}^n)_{\rm Weyl}$ where $m$ and $n$
are non-negative integers, and the subscript indicates that the product is
taken in the totally symmetric ordering.  The Hamiltonian is a linear
combination $\hat{H}=\frac{1}{2}(\hat{T}_{2,0}/m+m\omega^2\hat{T}_{0,2})$ of
$\hat{T}_{2,0}$ and $\hat{T}_{0,2}$, and therefore (\ref{eigen}) contains
products of the form $\hat{T}_{m,n} \hat{T}_{m',n'}$. Using the basic
commutation relation of $\hat{q}$ and $\hat{p}$, such products can always be
rewritten as sums over individual $\hat{T}_{m'',n''}$ of order $m+n+m'+n'$ or
less, as derived explicitly in \cite{OpDiff}. The condition (\ref{eigen}) is
therefore equivalent to a recurrence relation for
$\langle\hat{T}_{m,n}\rangle_E$ which is shown and discussed in more detail in
our supplementary material. (This material also uses an algebraic notion of
states \cite{LocalQuant} and makes contact with effective constraints
\cite{EffCons,EffConsRel}.)

In addition to higher-order moments $\langle\hat{T}_{m,n}\rangle_E$ of an
eigenstate, we have higher-order uncertainty relations. They can be obtained
just like Heisenberg's version, by applying the textbook derivation to integer
powers of $\hat{q}$ and $\hat{p}$ or their products instead of just $\hat{q}$
and $\hat{p}$. A systematic procedure to organize these higher-order, or
generalized, uncertainty relations has been given in
\cite{PositiveWigner,MomentsUncert}. For our purposes, a subset of these
relations is sufficient, which can be constructed as follows: We define
$\hat{\xi}_J$ as the $2J$-dimensional column vector consisting of all
$\hat{T}_{m,0}$ and $\hat{T}_{m-1,1}$ up to order $m=2J$, where $J$ is an
integer or half-integer. According to the generalized uncertainty principle,
the matrix $M_J = \langle\hat{\xi}_J\hat{\xi}_J^{\dagger}\rangle$ is
positive semi-definite for all $J$, where the expectation value is taken
element by element.  For $J=1/2$, we have Heisenberg's uncertainty principle
because a positive semi-definite matrix has a non-negative determinant.

As outlined in the supplementary material, positive semi-definiteness of $M_J$
can be reduced to the conditions
\begin{equation}\label{det}
\prod_{k=1}^n (E/\hbar\omega - \alpha_k)(E/\hbar\omega + \alpha_k)\geq 0 
\end{equation}
for all integer $n\geq 1$, where $\alpha_k = (2k-1)/2$ are the odd
half-integer multiples.  Considered as functions of $E$ for all $n$, these
expressions have nodes at $\hbar\omega\alpha_k$ up to some maximum $k$ that
depends on the particular value of $n$. Between nodes, the functions are
non-zero and alternate in sign. Moreover, sending $n$ to $n+1$ causes the
signs at fixed $E$ to alternate. This behavior combined with the
non-negativity of (\ref{det}) implies that the only allowable values for $E$
occur at the nodes. We can exclude negative values of $E$ because we have
already shown that $E\geq\frac{1}{2}\hbar\omega$.  Thus, the only possible
values for $E$ are such that $E/\hbar\omega =
\frac{1}{2},\frac{3}{2},\frac{5}{2}, \ldots$ in agreement with the well-known
eigenvalues of the harmonic oscillator.

Moreover, the arguments just given show that, for each eigenvalue
$E_n=(n-\frac{1}{2})\hbar\omega$, there is a generalized uncertainty relation
which restricts higher-order moments and is saturated by the corresponding
excited state with energy $E_n$. This result generalizes the well-known
statement that the ground state of the harmonic oscillator saturates
Heisenberg's uncertainty relation. Also note that our derivation, based on
expectation values, still applies if the state used is mixed, given by a
density matrix.  Since we obtain the usual energy spectrum of the harmonic
oscillator, it follows that mixed states do not to enlarge the spectrum.

As another consequence, we obtain the full energy spectrum of the harmonic
oscillator from the unfamiliar condition (\ref{eigen}) on energy
eigenvalues. This result serves as a proof of concept of the new algebraic
method introduced here, which we now apply to the Coulomb problem. We
will then be ready to generalize the results to non-associative quantum
mechanics, where the usual methods of computing eigenvalues are not
available. 

{\em Hydrogen:} The hydrogen atom has the Hamiltonian
$\hat{H}=\frac{1}{2}|\hat{p}|^2/m-\alpha \hat{r}^{-1}$ where
$|\hat{p}|^2=\hat{p}_x^2+\hat{p}_y^2+\hat{p}_z^2$ and
$\hat{r}^2=\hat{x}^2+\hat{y}^2+\hat{z}^2$. The position and momentum
components are subject to the basic commutation relations
$[\hat{x},\hat{p}_x]=[\hat{y},\hat{p}_y]=[\hat{z},\hat{p}_z]=i\hbar$.
For our purposes a different choice of
distinguished observables,
\begin{equation} \label{rPQ}
 \hat{r}\quad,\quad \hat{P}:=\hat{r}|\hat{p}|^2 \quad,\quad
 \hat{Q}:=\hat{x}\hat{p}_x+\hat{y}\hat{p}_y+\hat{z}\hat{p}_z \,,
\end{equation}
is more useful. Closely related variables have been used, quite differently,
to compute hydrogen spectra in deformation quantization
\cite{DefQuant2,DefQuantKepler,DefQuantHydro}. 

These operators have linear commutation relations
\begin{equation}
 [\hat{r},\hat{Q}] = i\hbar \hat{r}\quad,\quad
 [\hat{r},\hat{P}]=2i\hbar\hat{Q}\quad,\quad [\hat{Q},\hat{P}]=i\hbar\hat{P}\,,
\end{equation}
and there is a Casimir operator
\begin{equation}
 \hat{K}=\frac{1}{2}(\hat{r}\hat{P}+\hat{P}\hat{r})-\hat{Q}^2
\end{equation}
that commutes with $\hat{r}$, $\hat{P}$ and $\hat{Q}$. A direct calculation in
terms of the position and momentum components in (\ref{rPQ}) shows that
$\hat{K}$ is equal to the total angular momentum squared.  We should keep in
mind that not all the distinguished observables are self-adjoint. We do have
$\hat{r}^{\dagger}=\hat{r}$, but $\hat{Q}^{\dagger}=\hat{Q}-3i\hbar$ and
\begin{equation} \label{Pstar}
 \hat{P}^{\dagger}= \hat{P}-2i\hbar \hat{r}^{-1}\hat{Q}=\hat{P}-2i\hbar
 \hat{Q}\hat{r}^{-1}-2\hbar^2\hat{r}^{-1}\,. 
\end{equation}

As in our demonstration using the harmonic oscillator, we will be interested
in expectation values of monomials in $\hat{r}$, $\hat{P}$ and $\hat{Q}$
evaluated in eigenstates that obey (\ref{eigen}). We have another useful
relationship between certain expectation values given by the virial theorem:
\begin{equation} \label{Virial2}
\alpha\langle\hat{r}^{-1}\rangle_E=2E=-\frac{1}{m} \langle\hat{p}^2\rangle_E\,.
\end{equation}

The procedure used for the harmonic oscillator does not directly apply to the
Coulomb problem because the Hamiltonian is no longer quadratic, leading to
highly coupled recurrence relations.  We therefore reformulate the condition
(\ref{eigen}) in terms of a constraint linear in $\hat{P}$ and $\hat{r}$,
introducing
\begin{equation}
 \hat{C}_E=\hat{r}(\hat{H}-E)= \frac{1}{2m}\hat{P}-E\hat{r}-\alpha\,.
\end{equation}
The condition on the spectrum of $\hat{H}$ then takes the form
$\langle\hat{a}\hat{C}_E\rangle_E=0$ for all polynomials $\hat{a}$ in
$\hat{r}$, $\hat{r}^{-1}$, $\hat{P}$ and $\hat{Q}$. Unlike the Hamiltonian,
$\hat{C}_E$ is not self-adjoint. It is still useful to apply commutator
identities as in (\ref{aH}), but with a non-self-adjoint $\hat{C}_E$, there
are additional terms: In an eigenstate such that
$\langle\hat{a}\hat{C}_E\rangle_E=0$ and
$\langle\hat{a}^{\dagger}\hat{C}_E\rangle_E=0$,
\begin{equation}
 0=\langle\hat{a}\hat{C}_E\rangle_E-
 \overline{\langle\hat{a}^{\dagger}\hat{C}_E\rangle_E}=
   \langle(\hat{a}\hat{C}_E-\hat{C}_E^{\dagger}\hat{a}\rangle_E\,.
\end{equation}
With 
\begin{equation}
 \hat{C}_E^{\dagger}=\hat{C}_E-\frac{i\hbar}{m} \hat{r}^{-1}\hat{Q}=
\hat{C}_E-\frac{i\hbar}{m} \hat{Q}\hat{r}^{-1}-\frac{\hbar^2}{m}
\hat{r}^{-1}
\end{equation}
using (\ref{Pstar}), we have
\begin{equation} \label{aC}
 0=
 \frac{\langle[\hat{a},\hat{C}_E]\rangle_E}{i\hbar}+
 \frac{\langle\hat{Q}\hat{r}^{-1}\hat{a}\rangle_E}{m}-
 \frac{i\hbar\langle\hat{r}^{-1}\hat{a}\rangle_E}{m} \,.
\end{equation}

For $\hat{a}=\hat{Q}$,
\begin{equation}
 0=\frac{\langle\hat{P}\rangle_E}{2m} + E \langle\hat{r}\rangle_E+
\frac{\langle\hat{Q}^2\hat{r}^{-1}\rangle_E}{m}+ \frac{\hbar^2}{m}
\langle\hat{r}^{-1}\rangle_E\,. 
\end{equation}
If we replace $\hat{Q}^2$ using the Casimir operator $\hat{K}$, and
$\langle\hat{P}\rangle_E$ using $\langle\hat{C}_E\rangle_E=0$, we have
$0=3\alpha + 4E \langle\hat{r}\rangle_E- K_{\ell}
\langle\hat{r}^{-1}\rangle_E/m$.  The eigenvalues
$K_{\ell}=\ell(\ell+1)\hbar^2$ of $\hat{K}$ follow from angular-momentum
quantization, and $\langle\hat{r}^{-1}\rangle_E$ is related to $E$ by
(\ref{Virial2}). With these ingredients and similar calculations for
$\hat{a}=\hat{r}\hat{Q}$, we obtain
\begin{equation}
 \langle\hat{r}\rangle_E = \frac{1}{2}
 \frac{K_{\ell}}{m\alpha}-\frac{3}{4}\frac{\alpha}{E}\;,\;
 \langle\hat{r}^2\rangle_E= \frac{3}{4}\frac{K_{\ell}}{mE}+
 \frac{5}{8} \frac{\alpha^2}{E^2}- \frac{1}{4} \frac{\hbar^2}{mE}\,.
\end{equation}

In order to determine the allowed eigenvalues $E$, as before, we have to
impose uncertainty relations. We are interested here in the ground state, for
which we can focus on the lowest-order uncertainty relations, computed for our
non-canonical operators $\hat{r}$, $\hat{P}$ and $\hat{Q}$ using the
Cauchy--Schwarz inequality. There is only one non-trivial relation,
\begin{equation}  \label{rCUncert}
 (\Delta_E r)^2C^E_{\bar{Q}Q}\geq
 |C^E_{rQ}+\frac{1}{2}i\hbar\langle\hat{r}\rangle_E|^2\,, 
\end{equation}
with two covariances. Again using (\ref{aC}), we compute
$\langle\hat{Q}\rangle_E=\frac{1}{2}i\hbar$ using $\hat{a}=\hat{r}$,
$\langle\hat{r}\hat{Q}+\hat{Q}\hat{r}\rangle_E=i\hbar\langle\hat{r}\rangle_E$
using $\hat{a}=\hat{r}^2$. Finally, $\langle\hat{Q}^{\dagger}\hat{Q}\rangle_E=
\langle\hat{Q}^2\rangle_E-3i\hbar\langle\hat{Q}\rangle_E$ can be obtained
using $\hat{K}$.

Inserting all the required moments and factorizing the resulting polynomial in
$E$, (\ref{rCUncert}) gives the condition
\begin{widetext}
\begin{equation} \label{energy}
 \ell^2(\ell+1)^2(\ell^2+\ell-1)
 \frac{1}{E} \left(E+\frac{1}{2}
   \frac{m\alpha^2}{\hbar^2(\ell+1)^2}\right) 
 \left(E+\frac{1}{2} \frac{m\alpha^2}{\hbar^2\ell^2}\right)
 \left(E-\frac{1}{2}\frac{m\alpha^2}{\hbar^2(\ell^2+\ell-1)}\right)\geq 0\,.
\end{equation}
\end{widetext}
It is saturated for all energy eigenvalues with
maximal $\ell$, for which
\begin{equation} \label{El}
 E_{\ell+1}=-\frac{m\alpha^2}{2\hbar^2(\ell+1)^2}\,.
\end{equation}
Assuming the well-known degeneracy of the hydrogen
spectrum, we obtain the full set of bound-state energies. As in the example of
the harmonic oscillator, every eigenstate saturates an uncertainty relation,
in this case (\ref{rCUncert}).

{\em Non-associative hydrogen:} We are now in a position to derive our main
result. In the presence of a magnetic central charge, we cannot use canonical
momenta because they require a vector potential of the magnetic field
$\vec{B}$. Instead, we generate an algebra using kinematical electron momenta,
quantizing $p_i=m\dot{x}_i$. Their commutators are obtained by generalizing
the case in which there is a vector potential $\vec{A}$ depending on
$\vec{x}$, and canonical momenta are $\pi_i=p_i+eA_i$.  Therefore,
\begin{equation} \label{pp}
 [\hat{p}_j,\hat{p}_k]=
i \hbar e \left(\widehat{\frac{\partial A_k}{\partial x_j}}-
  \widehat{\frac{\partial  A_j}{\partial x_k}}\right)=
i\hbar e\sum_{l=1}^3\epsilon_{jkl}
 \hat{B}^l
\end{equation}
while $[\hat{x}_j,\hat{p}_k]=i\hbar\delta_{jk}$ is unchanged.

The final result depends only on $\vec{B}$ and therefore can be used to define
the commutators $[\hat{p}_j,\hat{p}_k]$ also if $\nabla\cdot\vec{B}\not=0$ in
the presence of magnetic charges.  A direct calculation shows that these
commutators then no longer obey the Jacobi identity:
\begin{eqnarray}
&& [[\hat{p}_x,\hat{p}_y],\hat{p}_z]+  [[\hat{p}_y,\hat{p}_z],\hat{p}_x]+
 [[\hat{p}_z,\hat{p}_x],\hat{p}_y]\nonumber \\
&=& i\hbar e \sum_{j=1}^3[\hat{B}^j,\hat{p}_j]= -\hbar^2e\;\widehat{{\rm
    div}\vec{B}}\not=0\,.
\end{eqnarray}
Even a single point-like monopole cannot be excised, as in Dirac's
construction, if we consider weak charges that do not obey the quantization
condition. However, a non-associative algebra generated by commuting
$\hat{x}_i$ and non-commuting $\hat{p}_j$, with standard commutators between
$\hat{x}_i$ and $\hat{p}_j$, is still meaningful \cite{Malcev,JackiwMon}.

Another direct calculation shows that the commutators of
$(\hat{r},\hat{Q},\hat{P})$ remain unchanged provided that
$\vec{r}\times\vec{B}=0$. This result, which relies on unexpected
cancellations of the extra terms in commutators implied by (\ref{pp}), is
crucial for the new application in this letter. In this case,
$\vec{B}=g(\vec{r})\vec{r}$. For a static magnetic field, we have
$\nabla\times\vec{B}=0$, which implies that $g(r)$ is spherically symmetric. A
monopole density $\nabla\cdot\vec{B}\not=0$ then requires that $g(r)= Q_{\rm
  m}(r)/(4\pi r^3)$ with the magnetic charge
\begin{equation}
 Q_{\rm m}(r)=4\pi \int \nabla\cdot\vec{B}(r) r^2{\rm d}r
\end{equation}
enclosed in a sphere of radius $r$. For a single monopole at $r=0$,
$g(r)=g$ is constant.

The virial theorem relies only on algebraic properties and remains valid. With
monopole commutators for momentum components, however, the modified angular
momentum $\hat{\vec{L}}{}'=\hat{\vec{L}}+eg\hat{\vec{r}}/\hat{r}$, not
$\hat{\vec{L}}$ itself, satisfies the usual commutators of angular momentum
\cite{MonopoleAngMom,MagneticCharge}. The Casimir of the algebra generated by
$(\hat{r},\hat{Q},\hat{P})$ is still equal to $\hat{K}=\hat{\vec{L}}{}^2$, but
in terms of the modified angular momentum it has an extra term:
\begin{equation} \label{KL}
 \hat{K}=\hat{\vec{L}}{}^2=\hat{\vec{L}}{}'{}^2- e^2g^2\,.
\end{equation}
For a single monopole at the center, the spectrum of $\hat{K}$ has a simple
shift compared with the standard spectrum of $\hat{L}^2$, which is known to
break the $\ell$-degeneracy of the hydrogen spectrum
\cite{MonopoleHydro}. Moreover, the allowed values of $\ell$ are restricted
for non-zero $g$ because $\hat{K}$, by definition, is positive, and so
must be its eigenvalues. Therefore, $\ell=0$ is not possible for
$g\not=0$, and larger $\ell$ may be ruled out as well for strong magnetic
charges. 

We will focus now on the range of weak magnetic charges given by
\begin{equation}
 0< \frac{eg}{\hbar}=N < \frac{1}{2}\,.
\end{equation}
None of these values could be modeled by a Dirac monopole
(they would not correspond to single-valued wave functions),
but they can be
considered if quantum mechanics is non-associative. Since the algebraic
relations used to derive (\ref{energy}) are still applicable, we obtain
conditions on the energy spectrum. The only difference is that the eigenvalues
of $\hat{K}$ are now given by $K_{\ell}=\ell(\ell+1)\hbar^2-e^2g^2$, which can
be taken into account by replacing $\ell$ in (\ref{energy}) with
\begin{equation} \label{ltilde}
 \tilde{\ell}=\sqrt{\left(\ell+\frac{1}{2}\right)^2-
\frac{e^2g^2}{\hbar^2}}-\frac{1}{2}\,.
\end{equation}
For quantized magnetic charges, the corresponding eigenvalues for which the
first parenthesis in (\ref{energy}) is zero are indeed
included in the spectrum found in \cite{MonopoleHydro}, but they no longer
constitute the full spectrum.

For weak magnetic charges, positivity of $\hat{K}$ requires that the smallest
possible $\ell$ is $\ell=1/2$, which we use for the ground state. The
corresponding $\tilde{\ell}$ is equal to
\begin{equation}
 \tilde{\ell}=\sqrt{1-N^2}-\frac{1}{2}
\end{equation}
and lies in the range $\frac{1}{2}(\sqrt{3}-1)<
\tilde{\ell}<\frac{1}{2}$. This range does not come close to the integer values
$0$ or $1$ which would amount to standard hydrogen eigenvalues. Therefore,
even for weak magnetic monopoles the energy spectrum of hydrogen is strongly
modified. The ground-state energy is discontinuous in the central magnetic
charge as a consequence of the positivity condition $K\geq 0$, which is the
reason why even a small magnetic charge is not a simple perturbation of the
usual hydrogen spectrum. 

This result would seem to rule out any non-zero magnetic charge of the
proton. However, from a purely experimental perspective, the smallest
eigenvalue of the total angular momentum, used in our evaluation of $K\geq 0$,
is zero only within some uncertainty. The angular momentum spectrum is very
basic and hard to modify. For instance, the conservation law and its role
played in parity considerations implies that, for a single component, it has
the form of a ladder centered around zero. It is, however, conceivable that
its values are washed out to within some $\delta L^2$.  To estimate this
quantity, we are not restricted to hydrogen-like systems because all energy
levels depend in some way on the eigenvalues of $\hat{L}^2$. The best relative
precision, of about $5\cdot 10^{-19}$, is obtained for spectral lines used in
atomic clocks \cite{LatticeClock}.  In SI units, a non-zero upper bound
\begin{equation}
 g\leq \frac{4\pi \epsilon_0\sqrt{\delta L^2}c^2}{e}\approx 4.7\cdot 10^{-18}{\rm
   Am}= 1.4 \cdot 10^{-9} g_{{\rm Dirac}}
\end{equation}
then follows from $K\geq 0$ and (\ref{KL}), where $g_{{\rm Dirac}}$ is the
smallest magnetic charge allowed by Dirac. 

For the proton, this bound is not as strong as existing ones
\cite{MagneticChargeBound,MagneticChargeProtonNeutron}. However, the bounds in
\cite{MagneticChargeBound,MagneticChargeProtonNeutron} are obtained by
limiting the total magnetic charge of a macroscopic object, adding the
individual charges of all electrons or nucleons. Our bound is obtained
directly for a single proton. Moreover, the magnetic charge of the muon is
more difficult to bound \cite{MagneticChargeProtonNeutron}. Our bound, on the
other hand, also applies to a muon as the nucleus of muonium, and to
antimatter such as the antiproton in antihydrogen \cite{AntiH,AntiH2} or the
positron in positronium \cite{Positronium}.

If we directly apply hydrogen or muonium spectroscopy, with accuracies of
$\Delta E/E\approx 4.5\cdot 10^{-15}$ \cite{HydrogenSpec} and about $10^{-9}$
\cite{MuoniumSpec}, respectively, we obtain weaker bounds: $g_{\rm proton}\leq
9.5\cdot 10^{-8}g_{{\rm Dirac}}$ and $g_{\rm muon}\leq4.5\cdot 10^{-5}g_{{\rm Dirac}}$.

\noindent {\em Acknowledgements:}
This work was supported in part by NSF grant PHY-1607414.  This research was
supported in part by the Ministry of Science, ICT \& Future Planning,
Gyeongsangbuk-do and Pohang City and the National Research Foundation of Korea
(Grant No.: 2018R1D1A1B07049126).

\section*{Supplementary material}

For an application to non-associative quantum mechanics, we present here
details of the systems considered in the letter in algebraic notation.

{\em Harmonic oscillator:} The quantum harmonic oscillator is defined by an
algebra generated by two distinguished observables $\hat{\tilde{q}}$ and
$\hat{\tilde{p}}$ with $[\hat{\tilde{q}},\hat{\tilde{p}}]=i$, containing the
Hamiltonian
$\hat{H}=\frac{1}{2}\hbar\omega(\hat{\tilde{p}}^2+\hat{\tilde{q}}^2)$. Compared
with the usual expressions, we have applied a linear transformation
$\hat{\tilde{q}}=\sqrt{m\omega/\hbar}\,\hat{q}$,
$\hat{\tilde{p}}=\hat{p}/\sqrt{m\omega\hbar}$ to simplify the original
Hamiltonian $\frac{1}{2}(\hat{p}^2/m+m\omega^2\hat{q}^2)$. The
same transformation removes $\hbar$ from the commutator.

The algebra is equipped with an adjointness relation such that
$\hat{\tilde{q}}^{\dagger}=\hat{\tilde{q}}$ and
$\hat{\tilde{p}}^{\dagger}=\hat{\tilde{p}}$, which implies
$\hat{H}^{\dagger}=\hat{H}$. A quantum state, in algebraic terms
\cite{LocalQuant}, is a positive linear map $\Omega$ from the algebra to the
complex numbers, such that $\Omega(\hat{a}^{\dagger}\hat{a})\geq 0$ for all
algebra elements $\hat{a}$. The positivity condition imposes uncertainty
relations, and it also implies that the state is normalized,
$\Omega(\hat{1})=1$. If the algebra is represented on a Hilbert space, any
wave function $|\psi\rangle$ or density matrix $\hat{\rho}$ defines such a map
by $\Omega_{|\psi\rangle}(\hat{a})=\langle\psi|\hat{a}|\psi\rangle$ or
$\Omega_{\hat{\rho}}(\hat{a})={\rm tr}(\hat{\rho}\hat{a})$, respectively. The
algebraic notion, however, can also be used in cases without a Hilbert space,
as in non-associative quantum mechanics. As in the letter, we define an energy
eigenstate $\Omega_E$ of $\hat{H}$ with eigenvalue $E$ by the condition
\begin{equation} \label{eigen}
 \Omega_E(\hat{a}(\hat{H}-E))=0
\end{equation}
for all algebra elements $\hat{a}$, or all polynomials in $\hat{\tilde{q}}$
and $\hat{\tilde{p}}$.

Following \cite{PositiveWigner,MomentsUncert}, we define the algebra elements
$\hat{T}_{m,n} := (\hat{\tilde{q}}^m\hat{\tilde{p}}^n)_{\rm Weyl}$ where $m$
and $n$ are non-negative integers, and the subscript indicates that the
product is taken in the totally symmetric ordering.  Through the basic
commutation relation, products of the form $\hat{T}_{m,n} \hat{T}_{m',n'}$ can
always be rewritten as sums over individual $\hat{T}_{m'',n''}$ of order
$m+n+m'+n'$ or less; see \cite{OpDiff} for an explicit statement of the
relevant reordering identity.  For an eigenstate of $\hat{H}$ with eigenvalue
$E$, we have $\Omega_E(\hat{T}_{m,n}(\hat{H} - E)) = 0$ for all $m,n\geq
0$. Since $\hat{H}=\frac{1}{2}\hbar\omega(\hat{T}_{2,0}+\hat{T}_{0,2})$, this
equation provides recurrence relations for $\Omega_E(\hat{T}_{m,n})$:
\begin{eqnarray}
&&\Omega_E(\hat{T}_{m+2,n}) + \Omega_E(\hat{T}_{m,n+2}) =\frac{n(n-1)}{4}
\Omega_E(\hat{T}_{m,n-2})\nonumber\\
&& \qquad+ \frac{m(m-1)}{4} 
\Omega_E(\hat{T}_{m-2,n})+ 2\frac{E}{\hbar\omega}
\Omega_E(\hat{T}_{m,n}) \label{Rec1} 
\end{eqnarray}
and
\begin{equation}
n\Omega_E(\hat{T}_{m+1,n-1}) = m\Omega_E(\hat{T}_{m-1,n+1})
\end{equation}
hold for all $m,n\ge0$.

From the second relation, starting with $m=0$ or $n=0$, we see that the
moments are zero unless both $m$ and $n$ are even.  Only moments of the form
$\Omega_E(\hat{T}_{2m,2n})$ are therefore non-zero. Since (\ref{Rec1}) is
symmetric under switching $m$ and $n$, $\Omega_E(\hat{T}_{2m,2n})$ is a
function only of $m+n$. Starting with the initial condition
$\Omega_E(\hat{T}_{0,0})=1$ using the normalization condition of states,
(\ref{Rec1}) implies
\begin{equation} \label{m1}
\Omega_E(\hat{T}_{2m,2n})= \frac{(2m)!(2n)!(m+n)!}{m!n!(2m+2n)!} a^E_{m+n}\,,
\end{equation}
where the $a^E_\ell$ are determined by the recurrence relation
\begin{equation} \label{m2}
a^E_{\ell+1} = \frac{E}{\hbar\omega}\frac{(2\ell+1)}{\ell + 1}a^E_\ell +
\frac{(2\ell+1)(2\ell)(2\ell-1)}{8(\ell + 1)}a^E_{\ell-1}
\end{equation}
in a single independent variable, $\ell$, with initial conditions $a^E_0 = 1$
and $a^E_1=E/\hbar\omega$ which follow from normalization and our
leading-order results presented in the letter, respectively.  Equations
(\ref{m1}) and (\ref{m2}) determine all orders of moments in terms of
$E/\hbar\omega$. It follows that $a^E_{\ell}$ is a polynomial in
$E/\hbar\omega$ of degree $\ell$, with only even terms for $\ell$ even and
only odd terms for $\ell$ odd.

So far, there is no restriction on $E$ because we have not yet imposed the
positivity condition on $\Omega_E$, or uncertainty relations.  Positivity can
efficiently be formulated using the $(J+1)(2J+1)$-dimensional column vector,
$\hat{\bar{\xi}}_J$, consisting of all $\hat{T}_{m,n}$ up to order $m+n=2J$,
where $J$ is an integer or half-integer.  Again following
\cite{PositiveWigner,MomentsUncert}, if a linear map $\Omega_E$ is positive,
the square matrix
\begin{equation}\label{uncertainty_princip}
\bar{M}_J = \Omega_E(\hat{\bar{\xi}}_J\hat{\bar{\xi}}_J^{\dagger}) \ge 0
\end{equation}
is positive semi-definite, where $\Omega_E$ is applied element by element.

It turns out that the positivity condition on $\bar{M}_J$ is redundant as far
as computing the spectrum of $\hat{H}$ is concerned.  The condition $\bar{M}_J
\ge 0$ implies that $M_J \ge 0$, where $M_J$ is a matrix formed by deleting
from $\bar{M}_J$ any number of rows and their corresponding
columns. Equivalently, $M_J$ is the matrix formed by deleting entries from
$\hat{\bar{\xi}}_J$ to form a new vector $\hat{\xi}_J$ and then taking
\begin{equation}
 M_J = \Omega_E(\hat{\xi}_J\hat{\xi}_J{}^{\dagger})\,.
\end{equation} 

In particular, consider the matrix $M_J$ formed by taking $\hat{\xi}_J$ to
contain only operators of the form $\hat{T}_{m,0}$ and $\hat{T}_{m-1,1}$.
The recurrence relation (\ref{m2}) implies relations between some of the
components of $M_J$, which can be exploited to bring
this matrix to block diagonal form, $(A_0,\ldots,A_{2J})$, with
$2\times 2$-matrices $A_n$. Then, $M_J \ge 0$ implies that
\begin{equation} \label{An}
A_n \ge 0 \hspace{3pt} \textrm{ for all } n \ge 0.
\end{equation}
For a fixed $n$, this is a constraint involving moments up to order $2n$,
which can in turn be written in terms of $E/\hbar\omega$ using (\ref{m1}) and
(\ref{m2}). Evaluating the determinants of $A_n$, 
\begin{equation} 
\det A_n(E) = \frac{1}{4^{n-1}} \prod_{k=1}^n (E/\hbar\omega -
\alpha_k)(E/\hbar\omega + \alpha_k)\geq 0
\end{equation}
with $\alpha_k = (2k-1)/2$, we obtain the energy spectrum as shown in the
letter.

{\em Hydrogen:} In the letter, we have shown the relevant details for the
Coulomb problem. Here, we give an alternative demonstration of the commutator
relationship used to compute moments of eigenstates. The derivation here
illustrates how our method in this example is related to quantum constrained
systems, which have been analyzed in terms of moments in
\cite{EffCons,EffConsRel}. These methods may be useful for further
derivations.

If $\Omega_E(\hat{a}\hat{C}_E)=0$,
$\Omega_E(\hat{a}F_E(\epsilon))=\Omega_E(\hat{a})$ with
$\hat{F}_E(\epsilon)=\exp(-i\epsilon\hat{C}_E/\hbar)$, for all real
$\epsilon$. The state $\Omega_E$ is therefore invariant under the flow
$\Omega_E(\hat{O})[\epsilon]=\Omega_E(\hat{F}_E(\epsilon)^{\dagger}
\hat{O}\hat{F}_E(\epsilon))$ generated by $\hat{C}_E$, akin to the evolution
generated by a Hamiltonian $\hat{H}$. However, since $\hat{C}_E$ is not
self-adjoint, $\hat{F}_E(\epsilon)$ is not unitary. Moreover, the
infinitesimal flow ${\rm d}\Omega_E(\hat{O})[\epsilon]/{\rm d}\epsilon$ is not
equal to the commutator $[\hat{O},\hat{C}_E]$, evaluated in $\Omega_E$ and
divided by $i\hbar$, but is given by
\begin{equation}
 \frac{{\rm d}\Omega_E(\hat{O})[\epsilon]}{{\rm d}\epsilon} =
 \frac{\Omega_E(\hat{O}\hat{C}_E-\hat{C}_E^{\dagger}\hat{O})[\epsilon]}{i\hbar}\,.
\end{equation}
With 
\begin{equation}
 \hat{C}_E^{\dagger}=\hat{C}_E-\frac{i\hbar}{m} \hat{r}^{-1}\hat{Q}=
\hat{C}_E-\frac{i\hbar}{m} \hat{Q}\hat{r}^{-1}-\frac{\hbar^2}{m}
\hat{r}^{-1}\,,
\end{equation}
we have
\begin{eqnarray}
 \frac{{\rm d}\Omega_E(\hat{O})[\epsilon]}{{\rm d}\epsilon} &=&
 \frac{\Omega_E([\hat{O},\hat{C}_E])[\epsilon]}{i\hbar}+
 \frac{\Omega_E(\hat{Q}\hat{r}^{-1}\hat{O})[\epsilon]}{m}\nonumber\\
&&-
 \frac{i\hbar\Omega_E(\hat{r}^{-1}\hat{O})[\epsilon]}{m} \,.
\end{eqnarray}
Invariance of $\Omega_E(\hat{Q})[\epsilon]$ then implies
\begin{eqnarray}
 0&=&\frac{{\rm d}\Omega_E(\hat{Q})}{{\rm
  d}\epsilon}\\
&=&\frac{\Omega_E(\hat{P})}{2m} + E \Omega_E(\hat{r})+
\frac{\Omega_E(\hat{Q}^2\hat{r}^{-1})}{m}+ \frac{\hbar^2}{m}
\Omega_E(\hat{r}^{-1}) \nonumber
\end{eqnarray}
as evaluated in the letter.

We have used angular-momentum eigenvalues in order to determine eigenvalues of
the Casimir operator $\hat{K}$. In the non-associative model, in particular,
we applied the standard eigenvalues to the modified angular momentum
$\hat{\vec{L}}'$ which obeys the same commutator algebra as the standard
$\hat{\vec{L}}$. This step might be questioned becaue we do not have a
Hilbert-space representation of a non-associative algebra, and therefore it is
not obvious how to apply the standard derivation, for instance using ladder
operators $\hat{L}'_{\pm}$. However, a brief argument shows that the usual
eigenvalues are still correct: An algebraic state $\omega$, that is, a mapping
from the algebra of observables to the complex numbers, induces a state on any
unital subalgebra, such as the algebra spanned by the components of
$\hat{\vec{L}}'$ together with the identity operator. Since this subalgebra is
associative even if the full algebra is non-associative, we can use the
Gelfand--Naimark--Segal (GNS) construction \cite{LocalQuant} to induce a
Hilbert-space representation. In this representation, the standard derivation
of eigenvalues applies, such that the modified angular momentum
$\hat{\vec{L}}'$ of a non-associative theory does not have modified
eigenvalues.

\end{document}